\documentclass[12pt,epsf,epsfig,psfig]{article}
\usepackage{graphicx}
\usepackage{epsfig}
\def\J{$J/\psi$}

\def\X{$\chi_c$}

\def\P{$\psi'$}

\def\Q{Q{\bar Q}}
\def\e{\epsilon}

\def\NP{{ Nucl.\ Phys.\ }}
\def\PL{{ Phys.\ Lett.\ }}
\def\PR{{ Phys.\ Rev.\ }}

\def\PRL{{ Phys.\ Rev.\ Lett.\ }}

\def\ZP{{ Z.\ Phys.\ }}
\def\EPJ{{Eur.\ Phys.\ J.\ }}

\def\be{\begin{equation}}
\def\ee{\end{equation}}
\def\lsim{\raise0.3ex\hbox{$<$\kern-0.75em\raise-1.1ex\hbox{$\sim$}}}
\def\gsim{\raise0.3ex\hbox{$>$\kern-0.75em\raise-1.1ex\hbox{$\sim$}}}

\begin{document}

\thispagestyle{empty}
January 2011 \hfill BI-TP 2011/02
%19.1.11

~~~~\vskip 0.5cm

\centerline{\LARGE \bf The Quark-Gluon Plasma$^*$}

\bigskip

\centerline{\Large \bf A Short Introduction}

\vskip1cm

\centerline{\large \bf Helmut Satz} 

\bigskip

\centerline{Fakult\"at f\"ur Physik, Universit\"at Bielefeld, Germany}

\vskip1cm

\section{States of Matter}

The quark-gluon plasma is a state of strongly interacting matter, in which
the quarks and gluons, which make up hadrons, are not longer confined to
color-neutral entities of hadronic size. What does that mean?

\medskip
 
\begin{figure}[htb]
\centerline{\epsfig{file=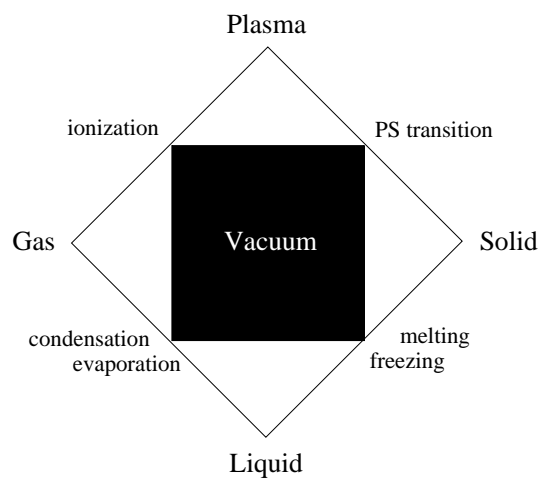,width=7cm}}
\caption{Classical states of matter and transitions between them}
\label{phases}
\end{figure}

Matter, in statistical mechanics, is a system of many constituents in 
local thermal equilibrium -- i.e., a system whose average properties are
specified by a few global observables (temperature, energy density, net 
``charge''). For different values of these observables, the system may
exhibit fundamentally different average properties, and so there exist 
different states of matter, with ``phase'' transitions occurring when the
system changes from one state to the other. The classical pattern,
already proposed in both Greek and Hindu natural philosophy, has
the form shown in Fig.\ \ref{phases}. To 
four fundamental states:
solid, liquid, gas and plasma, the vacuum is added as fifth element
(``quintessence''), providing the space in which matter exists. What are
the states of matter in the sub-atomic world of strong interactions?

\vfill

\hrule width4cm

\vskip0.2cm

{* Student Day Lecture, {\sl 6th International Conference on
Physics and Astrophysics of Quark Gluon Plasma}, 
Dec.\ 5 - 10, 2010, Goa/India}

\newpage

To get a first idea, let us begin with a very simple
picture. If nucleons, with their given spatial extension, were both
elementary and incompressible, then a state of close packing would
constitute the high density limit of matter. 
%(Fig.\ \ref{states}). 
If, on the other hand, nucleons are really composite -- bound states
of point-like quarks --, then with increasing density they will start 
to overlap, until eventually we reach a state in which each quark finds
within its immediate vicinity a considerable number of other quarks.
It has no way to identify which of these had been its partners in a
specific nucleon at some previous state of lower density. Beyond a
certain point, the concept of a hadron thus loses its meaning, 
and we are quite naturally led from nuclear matter to a system whose 
basic constituents are unbound quarks.

\medskip

More specifically, in confined matter the constituents are color-neutral
quark-antiquark or three-quark states of hadronic size (radius $\sim$ 1 fm).
The quarks inside a hadron polarize the surrounding gluonic medium; the 
resulting gluon cloud around each quark provides it with a 
dynamically generated effective mass of about 300 MeV. In an ideal version
of QCD, with massless quarks in the Lagrangian, this corresponds to 
spontaneous chiral symmetry breaking.   

\medskip

Confined hadronic matter exists in two distinct forms. At vanishing or
low baryon number density, it consists largely of mesons, since the 
higher baryon mass reduces, through the Boltzmann factor, the baryonic
thermodynamic weight. The interaction between mesons as well as that 
between mesons and baryons is resonance-dominated, i.e., it consists
essentially of the abundant formation of multi-meson and meson-baryon 
resonances. Mesons appear to allow arbitrary overlap,
so all (light quark) hadron states have the same characteristic
size, with a radius of about 1 fm, independent of their mass. Nucleons,
on the other hand, experience a short range repulsion in addition to
a longer range attraction. The latter, the nuclear force, binds nucleons
to nuclei, while the former leads to a nuclear volume linear in the 
total number of nucleons. This implies that nucleons, also with
a hadronic radius of about 1 fm, have an effective hard core of about
half that size. Both these forces are of non-resonant nature, so that
the interaction in baryonic matter at low temperature and high density 
is thus quite different from that of mesonic matter. Nevertheless, in
each case we have with increasing density, be it through ``heating''
or ``compression'', a cluster formation which eventually leads to more
quarks per hadronic volume than meaningful for a partioning into 
color-neutral hadrons. In other words, increasing of the temperature
$T$ or the baryochemical potential $\mu$ results eventually in color
deconfinement. What happens in this transition?

\medskip

On one hand, the color-neutral states are dissolved, producing a medium
of color-charged constituents. The deconfinement transition is thus the
QCD counterpart of the insulator-conductor transition of atomic physics.
In addition, a sufficient increase of the temperature results eventually
in a ``melting'' of the gluon cloud which surrounds the quarks inside
a hadron. Hadronic matter thus shows two transitions, deconfinement and
chiral symmery restoration. Do these two phenomena necessariy coincide? 

\medskip

Rather general basic arguments \cite{B-C} show that they either occur
at the same point or if not, deconfinement precedes chiral symmetry
restoration. A simplistic justification is given by the fact that any
$r$-dependent confinement potential breaks chiral symmetry - a
reflection of the quark at the potential wall does not flip its spin.

\medskip
 
It is thus possible that quarks, when they become deconfined, still
maintain their effective mass up to some higher temperature or density.
Lattice calculations have shown that for vanishing baryon density, 
deconfinement and chiral symmetry restoration do in fact coincide,
indicating that the deconfinement temperature is sufficient to melt the
effective quark mass. For high baryon density at low temperature, this
seems not likely, allowing a medium of massive quarks as an additional 
state of strongly interacting matter, in addition to hadronic matter 
and the plasma of deconfined massless quarks and gluons \cite{CGS}.

\medskip

\begin{figure}[h]
\centerline{\psfig{file=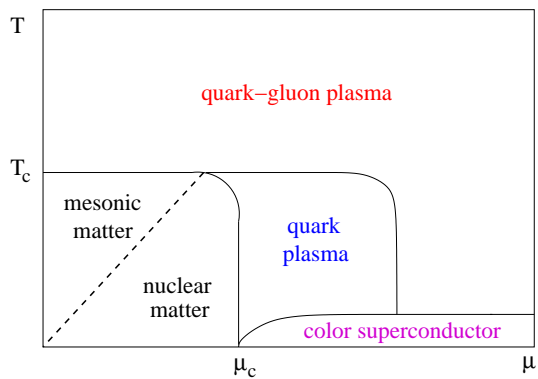,width=7cm}}
\caption{Speculative phase diagram of strongly interacting matter}
\label{phase-diagram}
\end{figure}

In any case, deconfinement does not result in a non-interacting medium;
in section 3, we shall return in detail to the strong interaction in the 
QGP in the region above the onset of deconfinement. Here we only note
that in particular the anti-triplet quark-quark interaction provides
an attractive force, making possible the existence of diquarks as
localized bound states. Such colored bosonic states can condense and
therefore form a color superconductor as yet another state of strongly 
interacting matter. Putting everything together gives us the 
phase diagram shown in Fig.\ \ref{phase-diagram}.  

\medskip

After this discussion, meant to show how the QGP fits into the general 
context of strong interaction thermodynamics, we shall now turn to its
study in the low baryon number regime.  Here extensive lattice calculations 
provide much information; moreover, this is also the region of relevance 
for RHIC and LHC experiments. 

\section{From Hadronic Matter to QGP}

\begin{figure}[htp]
\centerline{\psfig{file=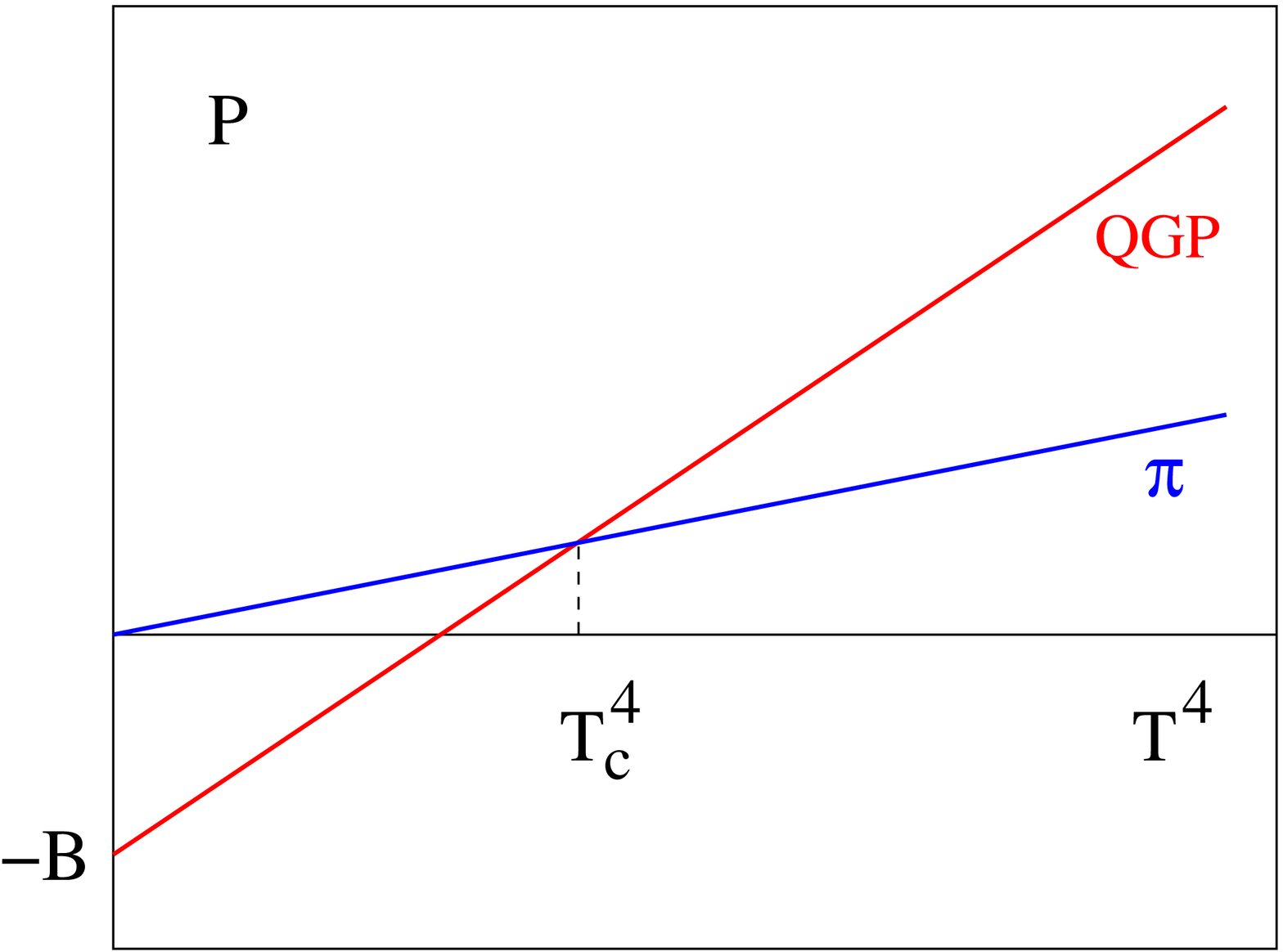,width=6cm} \hspace*{2cm}
\psfig{file=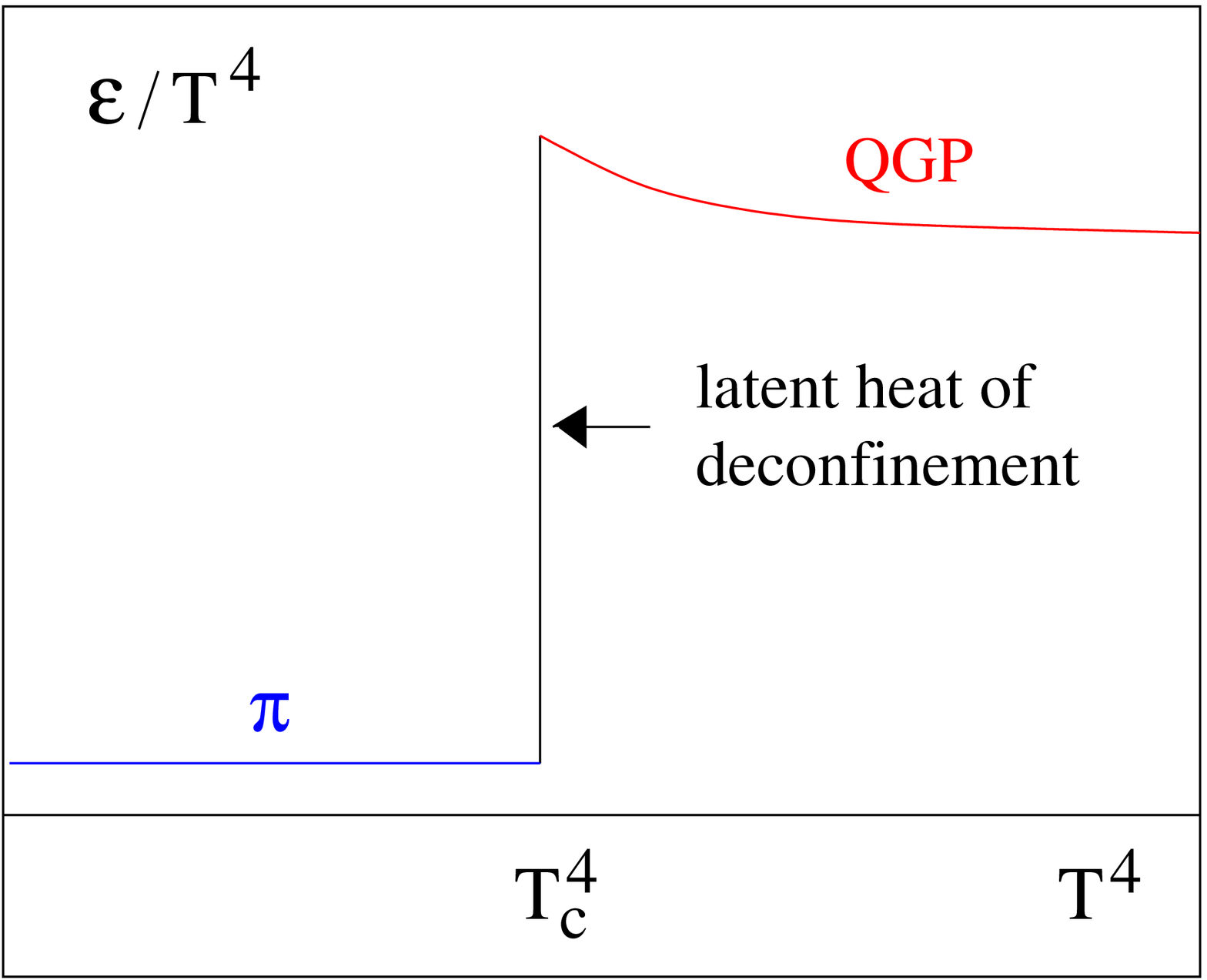,width=5.6cm}}
\caption{Pressures and energy densities for an ideal pion gas compared
to an ideal QGP}
\label{comp}
\end{figure}

The simplest form of confined matter is an ideal gas of massless pions, 
whose pressure is given by
\be
P_{\pi} = {\pi^2 \over 90}~3~ T^4 \simeq {1\over 3}~ T^4,
\label{piongas}
\ee
taking into account the three possible pion charge states. For deconfined
matter, we have as simplest case an ideal quark-gluon plasma (two 
massless quark flavors, two quark and two gluon spin orientations, 
$q$ and $\bar q$; three quark and eight gluon color degrees of freedom), 
giving
\be
P_{QGP} = {\pi^2 \over 90}~ \{~\!2\times 8 + {7 \over 8}~ [ 2 \times 2
 \times 2 \times 3 ]~\!\}~ T^4 - B  \simeq 4~T^4 - B
\label{qgp}
\ee
for the pressure, with $B$ denoting the ``bag pressure'' excerted by
the physical vacuum on the colored medium. The two forms of the pressure
are compared in Fig. \ref{comp} (left). 

\medskip

Since nature always chooses the state of highest pressure
(lowest free energy), this implies a phase transition from a pion gas
at low $T$ to a QGP at high $T$. The critical temperature is obtained
from $P_{\pi}=P_{QGP}$, leading to $T_c^4\simeq 0.3~B$ and hence
$T_c \simeq 150$ MeV, if we use $B^{1/4} \simeq 200$ MeV, as obtained
from quarkonium spectroscopy. The corresponding energy densities become
$\e_{\pi} \simeq T^4$ and $\e_{QGP}\simeq 12~T^4 + B$. leading to the
form shown in Fig.\ \ref{comp} (right). 

\medskip

The transition in this model is by construction of first order; at $T_c$, 
the energy density changes abruptly by the latent heat of deconfinement.
For an ideal gas of massless constituents, the energy density and pressure
are related by $\e=3P$; in such a conformal world, the temperature is
the only scale. In our case, the interaction measure
\be
\Delta \equiv {\e - 3P \over T^4} = {4B \over T^4}
\label{inter}
\ee
is definitely not zero; even in this simplistic model, the QGP is 
non-conformal (with $B$ as external scale) and strongly interacting 
for $T_c \leq T < 3-4~T_c$.

\medskip

To turn from model to theory, we want to derive the thermodynamics 
obtained with QCD as dynamical input. The only possible {\sl ab initio} 
calculations are based on the computer simulation \cite{Creutz}
of the lattice regularization of QCD \cite{Wilson}. We therefore summarize 
the main results thus obtained in finite temperature lattice QCD. 

\medskip

\begin{figure}[htb]
\vspace*{-0.5cm}
\centerline{\epsfig{file=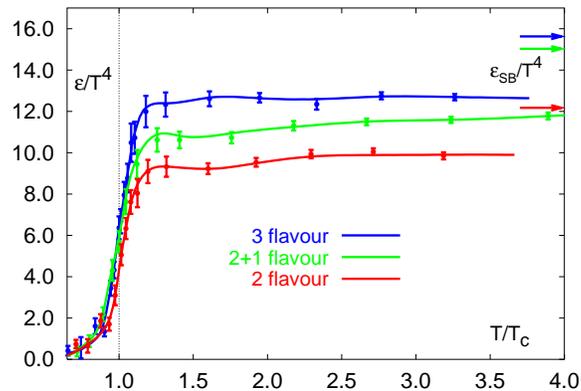,width=8cm}}
\caption{Lattice results for 
energy density vs.\ temperature in QCD thermodynamics \cite{KLP}}
\label{edens}
\end{figure}

In Fig.\ \ref{edens} we show the temperature behavior of the energy
density for the cases of 2, 2+1 and 3 light quark flavors \cite{KLP};
here 2+1 means one heavy
and two light quarks. The sudden jump corresponding to the latent heat
of deconfinement is quite evident. It is found to occur at a critical
temperature of about 160 - 180 MeV, and the energy density at that point
is around 0.5 to 1.0 GeV/fm$^3$. 

\medskip

To relate this ``jump'' more specifically to some form of critical
behavior, we consider the corresponding order parameters for deconfinement
and for chiral symmetry restoration. Such parameters signal the onset of
the new phase. 

\medskip

For deconfinement, the order parameter is given by the
average value of the Polyakov loop \cite{Larry},
\be
\langle L(T) \rangle \sim \exp\{-F_{Q\bar Q}(T)/T\},
\label{polya}
\ee
where $F_{Q\bar Q}(T)$ denotes the free energy of a static $\Q$ pair
at infinite separation. In a confining medium, this diverges 
($F_{Q\bar Q}(r,T)
 \sim \sigma r$), while in a deconfined medium, screening prevents 
communication between the $Q$ and the $\bar Q$ beyond a certain distance, 
so that $F_{Q\bar Q}$ remains finite. As a result,
\be 
L(T) 
~\left\{\matrix{=0 & T < T_L ~~~{\rm confinement}\cr
~&~\cr
\not= 0 &  ~~T > T_L~~~{ \rm deconfinement}\cr}
\right\}
\ee
defines the deconfinement temperature $T_L$. Actually, $\langle L(T) \rangle$
vanishes exactly for $T < T_L$ only in the case of infinitely heavy quarks;
for finite quark mass, the string binding $Q$ and $\bar Q$ breaks when the
potential surpasses the spontaneous pair formation threshold. As a result,
$\langle L(T) \rangle$ is very small but finite for $T < T_L$.

\medskip

The breaking of chiral symmetry is indicated by a finite value of the
chiral condensate $\chi(T) \equiv \langle \bar{\psi} \psi \rangle \sim M_q$, 
which measures the dynamically generated  (``constituent'') quark mass $M_q$, 
obtained for a Lagrangian with massless quarks. At high temperature, this 
mass melts, so that
\be
\chi(T) 
~\left\{\matrix{\not=0 & T<T_{\chi}~~ {\rm chiral~symmetry~broken}\cr
~&~\cr
=0 & ~T > T_{\chi} ~~{\rm chiral~symmetry~restored}\cr}
\right\}
\ee
defines the chiral symmetry restoration temperature $T_{\chi}$. Here
we have exact chiral symmetry only if the input quarks are massless;
for finite quark mass, the symmetry remains explicitly broken, so that then
$\chi(T)$ only becomes very small for $T > T_{\chi}$.

\medskip

Both $\langle L(T) \rangle$ and $\chi(T)$ have been studied extensively
in finite temperature lattice QCD at vanishing overall baryon number.
The corresponding susceptibilities (derivatives with respect to $T$)
peak sharply, defining $T_L$ and $T_{\chi}$, and within errors, the
two temperatures and hence the two phenomena (deconfinement and chiral
symmetry restoration) coincide. The critical temperature for the 
resulting transition 
from hadronic matter to QGP, for two light quark flavors, is thus determined 
as $T_c \simeq 175$ MeV.

\medskip

\begin{figure}[htb]
\centerline{\epsfig{file=scalingcernL.eps,width=7cm}\hskip1.5cm
\epsfig{file=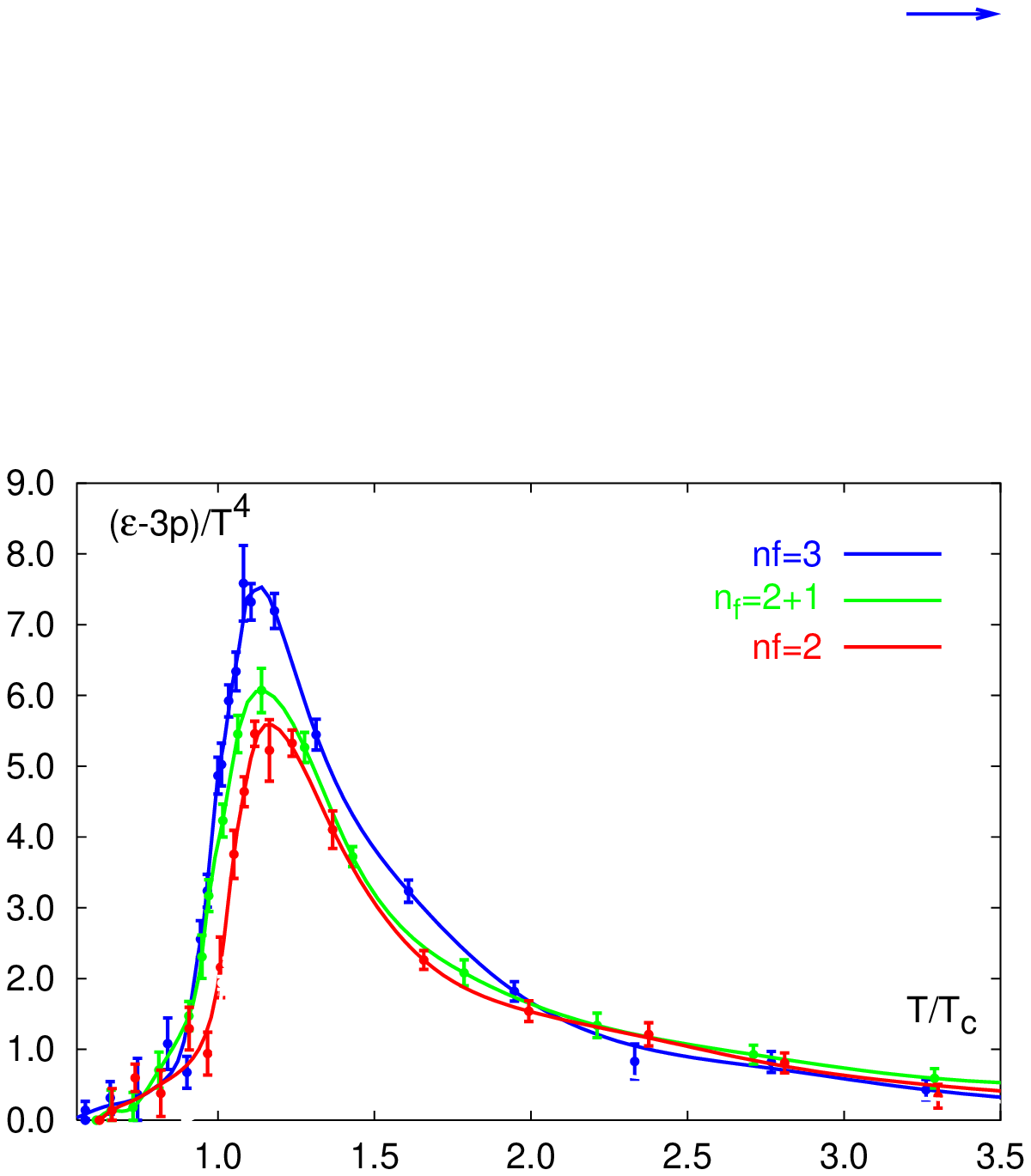,width=7.4cm}}
\label{delta}
\caption{Interaction measure in $SU(3)$ gauge theory (left)
\cite{DG} and full QCD (right) \cite{KLP}.}
\end{figure}

It was clear from the very first finite temperature lattice studies that
the deconfined medium in the region above $T_c$ is very strongly interacting 
and thus quite far from an ideal plasma \cite{inter-plasma}. This is best 
seen from the interaction measure (the trace of energy-momentum tensor), 
defined as
\be
\Delta = {\e - 3P \over T^4}.
\label{interaction}
\ee
For non-interacting massless constituents (the ``conformal'' limit),
$\Delta \equiv 0$, so that the temperature is the only scale. The
quarks and gluons of QCD are ideally massless, but the so-called
trace anomaly violates conformality and introduces a dimensional 
scale. The behavior of $\Delta(T)$ is shown in Fig.\ \ref{delta}
both in pure gauge theory for different color groups and for full
QCD with different flavor content. 

\medskip

\section{The Strongly Interacting QGP}

For sufficiently high temperature, asymptotic freedom is expected to
result in an ideal QGP. How high does $T$ have to be in order to allow
some form of a weak coupling expansion (perturbation theory) to describe
the approach to this limit?

\medskip 

Infrared divergences limit the perturbative expansion of QCD to $O(g^5)$
\cite{Linde}. The evaluation up to this order has been carried out 
\cite{Arnold} and is found to result in strongly oscillating and hence
non-convergent behavior for $T \leq 10~T_c$. 

\medskip

\begin{figure}[h]
\centerline{\psfig{file=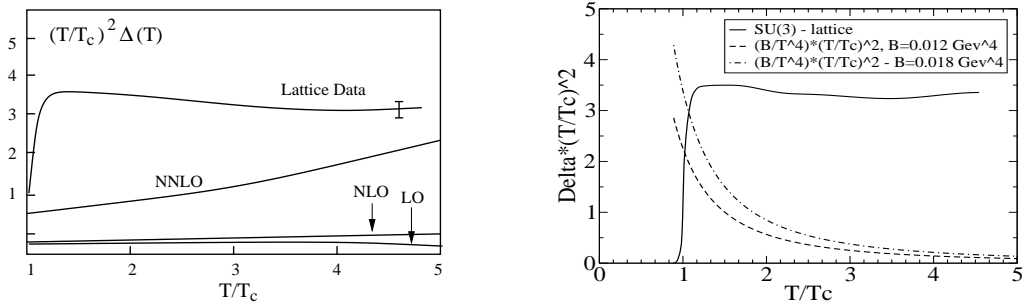,width=5.8cm,height=3.9cm}\hskip1.5cm
\epsfig{file=fitbag.eps,width=6.2cm,height=4cm}}
\caption{Interaction measure in $SU(3)$ gauge theory \cite{Boyd}, compared
to (left) HTL resummed perturbation theory \cite{HTL}, and (right)
bag model prediction \cite{CMS}.}
\label{HTL}
\end{figure}

This has led to considerable efforts to ``repair'' the difficulty,
either by introducing non-perturbative scale effects to allow a systematic
extension of perturbation theory beyond  $O(g^5)$ \cite{Laine}, or
by regrouping sets of Feynman diagrams to expand around a ground state
including screening effects (``resummed'' perturbation theory \cite{resum}, 
hard thermal loop approach \cite{HTL}). In both cases, however, such
weak-coupling methods cannot account for the behavior observed in lattice
studies. This holds in particular for $SU(3)$ gauge theory, where one
has results for the continuum limit \cite{Boyd}; see Fig.\ \ref{HTL} (left)
for a comparision to HTL results. It is obvious that no
weak-coupling approach can account for the critical behavior near $T_c$
(the dip of $\Delta(T)$ as $T\to T_c$); but also the behavior in the region
up to about 5 $T_c$, with $T^2 \Delta(T) \simeq {\rm const.}$, is not
reproduced by the weak logarithmic form of perturbative studies. 

\medskip

The simplest non-perturbative approach, using the bag model form discussed
above, does not fare any better. The bag pressure can be related to the
gluon condensate $G^2_0$ at $T=0$ \cite{Leutwyler}, and using numerical
estimates for the latter \cite{SVZ}, one obtains a $\Delta(T)$ vanishing
as $T^{-4}$ and thus much too fast; moreover, the critical dip near
$T_c$ is also here not given (see Fig.\ \ref{HTL} (right)). 

\medskip

We thus need to find a more detailed way to account for the non-perturbative
behavior observed in the region $T_c \leq T \leq 5~\!T_c$. One such
possibility is given by the quasi-particle approach, first applied to
$SU(2)$ gauge theory \cite{Golo93}. The basic idea is that in the deconfined
medium, gluons acquire an effective  mass by polarizing the colored medium 
around them. In the critical region, as $T \to T_c$ from above, the 
correlation length $\xi(T)$ increases strongly or diverges, so that the gluon
sees and polarizes more and more of the medium, increasing its mass. 
Outside the critical region, for $T \geq 1.3~\!T_c$, as $T$ increases
further, the correlation length decreases essentially as $T^{-1}$; now,
however, the energy density $\e(T)$ increases (as $T^4$), so that the effective
mass $M_g \sim \e \times \xi \sim T$ also grows. These two competing 
modes lead to a growth of $M_g$ for $T \to T_c$ determined by the
critical behavior and a conformal growth at high temperature. The dip 
in $M_g(T)$ around $T \sim 1.3~\!T_c$ defines the transition from critical
behavior to hot QGP. Applying this approach to the continuum limit of
$SU(3)$ gauge theory leads to excellent agreement, for he interaction measure
as well as for the energy density, in both the critical
region and in the hot QGP \cite{CMS}.

\medskip
   
We conclude that in the region of main interest to us, for $T_c \leq 
T \leq 3 - 5~\!T_c$, the QGP is strongly interacting. The interaction
provides the constituents with a temperature-dependent effective mass,
so that a considerable part of any energy input goes into mass rather
than into kinetic energy. As a result of this, the behavior of the 
equation of state is highly non-ideal and in fact not unlike that 
just below $T_c$, where abundant resonance formation restricts the
fraction of energy available for kinetic purposes.

\section{Probing the QGP}

At sufficiently high temperatures and/or densities, strongly interacting
matter thus becomes a plasma of deconfined, colored quarks and gluons.
How can we probe the properties of this medium and study its behavior
as function of temperature and density? This is a highly non-trivial
problem, and I want to outline here briefly three ways of addressing it.
We assume that we are given a macroscopic volume of deconfined strongly
interacting matter and want to determine its state for different
temperatures. This means that we shall consider equilibrium thermodynamics
only, leaving out all the phenomena (collisions effects, time dependence,
equilibration, flow) that make the analysis of actual nuclear collisions
data so complex.
 
\medskip

The given medium is by assumption hotter than its environment (the vacuum)
and hence emits radiation. An outside observer will detect the emission
of light hadrons; however, these cannot exist in the interior of the QGP 
and hence must be formed through hadronization at the cooler surface. 
Such radiation will therefore provide information about the hadronization
stage of the QGP, but not about the pre-hadronic state in the interior.
In the hot QGP itself, quark-gluon interactions and quark-antiquark 
annihilation produce real and virtual photons, respectively, and these
will leave the medium without further strong interaction. They can thus
provide information about the state of the medium when they were formed,  
i.e., about the hot QGP \cite{em-radiation}. The difficulty is that they 
can be formed at
all evolution stages of the medium, even in the hadronic phase, and so
one has to find a way to identify hot thermal electromagnetic radiation.
If this can be achieved, such radiation provides a thermometer for the
medium.

\medskip

Alternative tools are obtained by testing the medium with external
probes. In particular, we can study the effect of the medium on
quarkonia or on jets. Both will interact strongly in a deconfined medium
and less or not at all with hadronic matter; thus they can provide information 
on the temperature and/or density of the QGP.

\medskip

Quarkonia are bound states of heavy quark-antiquark pairs 
($c\bar c, b \bar b$). They
are much smaller than ``light'' hadroncs ($r_Q \ll r_h \sim 1$ fm) and
much more tightly bound, with binding energies up to 0.5 to 1.0 GeV. Therefore
they can survive in a QGP up to temperature above the deconfinement
point and ``melt'' only when the color screening radius has dropped to
quarkonium size \cite{MS}. Since the different quarkonium states have
different sizes and binding energies, since will lead to a ``sequential''
suppression of quarkonia: first, the larger and more loosely bound
excited states are dissolved, finally the small and tightly bound
ground states. For charmonia, this is illustrated in Fig.\ 7,
with \P~and \X~melting followed eventually by that of the \J. 
Such patterns can provide a spectral analysis of the
QGP, similar to that obtained for the sun by solar spectra
\cite{KS}.

\medskip

\begin{figure}[htb]
~\hskip2cm{\psfig{file=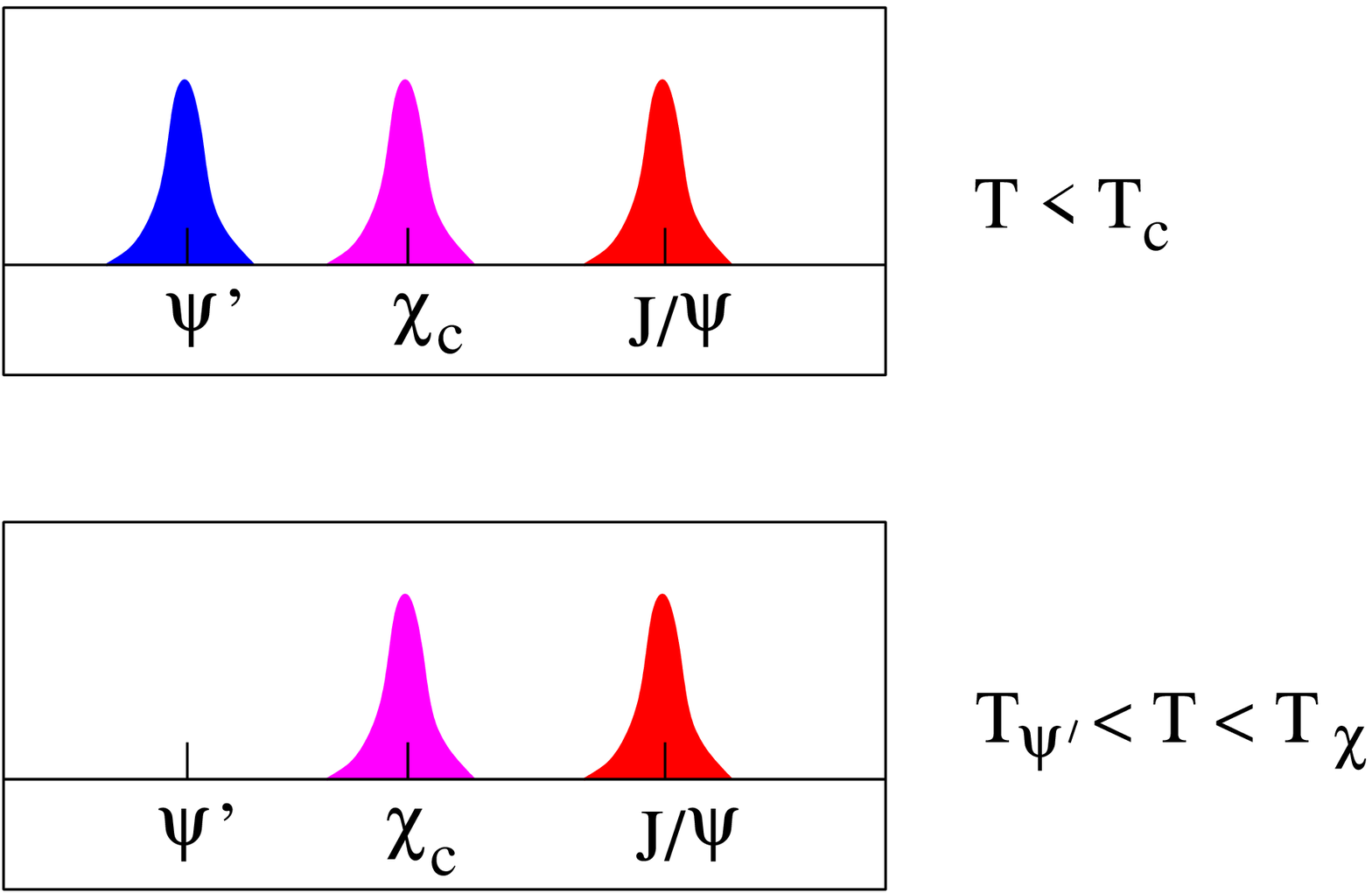,width=6cm}}
\end{figure}

\begin{figure}[htb]
\vspace*{-4.7cm}
\hfill{\psfig{file=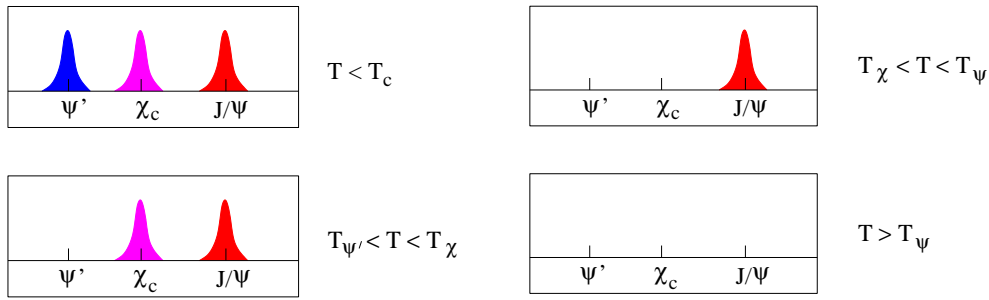,width=6.1cm}}~~~~~
\label{spectral}
\caption{The spectral analysis of the QGP through charmonium states}
\end{figure}

Jets are fast partons (quarks or gluons) passing through the medium.
They are colored and hence interact stronger with a QGP than with
color-neutral hadronic matter. A substantial attenuation (``quenching'')
of jets thus indicates the presence of a dense deconfined medium
\cite{Bj,Schiff}.

\medskip

We had called quarkonia and jets ``external'' probes. It is clear,
however, that they have to be produced in the same collision which
leads to the QGP candidate to be probed. They are, however, produced
through very early hard interactions, which take place before the
QGP is formed. We can then study the subsequent effect of the QGP
on their behavior. Moreover, their initial production is to a large
extent calculable by perturbative QCD, and it can be gauged in the
study of $pp$ and $pA$ collisions, which presumably do not produce
a QGP.

\section{Three Questions to the LHC}

The QGP predicted by statistical QCD is the ultimate state of matter 
to be studied in high energy nuclear collisions. This is a speculative
endeavor, since it is not clear to what extent such collisions can 
produce something to be called matter. We therefore close our survey
with three questions to the next generation of experiments which might 
help us in finding an answer to this fundamental enigma.

\medskip 

If an increase of collision energy indeed leads to the production of a 
hotter bubble of deconfined primordial matter, then this must expand more 
in order to reach the hadronization temperature, and hence the source size
for hadron emission must become larger. In particular, it is expected to
increase as a power of the hadron multiplicity, since this in turn grows 
with the initial energy density \cite{Stock}. So far, from AGS to RHIC, 
the source size for hadron emission, as determined by Hanbury-Brown--Twiss 
(HBT) methods \cite{HBT} used in astrophysics, has not shown a significant 
increase \cite{Beca}. 
This ``HBT-puzzle'' has been accounted for in terms of the relative role
of meson and baryon production \cite{Adamova02}, but at LHC energies, a 
clear increase of the source volume is predicted. Such an increase seems
necessary in a model-independent way, if the concept of hot primordial
fireball production in nuclear collisions is to make any sense.

\medskip

We had noted that momentum spectra for real and virtual photons can in 
principle provide an internal thermometer of the QGP, with
\be
(dN_{\gamma}/dk_T) \sim \exp\{-k_T/T\}
\ee 
A recent analysis of RHIC $Au\!-\!Au$ data at $\sqrt s=200$ GeV
\cite{phenix} has identified possible thermal photons,
seen in a transverse momentum window between pion decay and prompt 
photon spectra. 
%(Fig.\ \ref{photon}).
The corresponding temperature is with 
$T=221 \pm 19({\rm stat.}) \pm 19({\rm syst.})$ MeV above the
hadronization value of about 175 MeV. If such thermal photons are
indeed observable, the LHC should lead to much higher temperatures 
for electromagnetic radiation.

\medskip

\begin{figure}[htb]
\centerline{\psfig{file=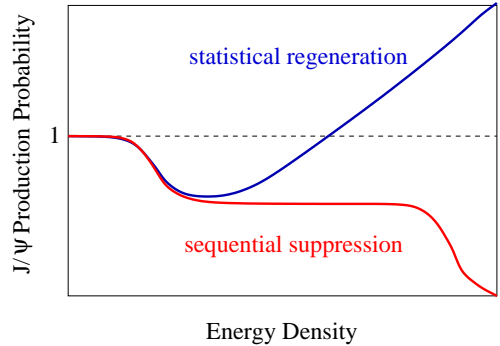,width=6.5cm}}
\caption{Sequential \J~suppression vs.\ statistical \J~regeneration}
\label{seq}
\end{figure}

The last question addresses quarkonium production in nuclear collisions
at the LHC. The \J~production rate in $Au-Au$ collisions at RHIC is 
compatible with that for central collisions at the SPS, once cold
nuclear matter effects are taken into account. The remaining survival
rate of about 50 \% is in accord with suppression of the higher
excited states (\P~and \X) and survival of the direct \J~\cite{KKS}.
The much higher energy density of the LHC should dissociate also the
latter, leading to complete \J~suppression (modulo $B$ decay and corona
production). The expected survival pattern is illustrated in Fig. \ref{seq}.

\medskip

Here, however, an alternative scenario has been proposed \cite{regen}
and much discussed. Charm production in nuclear collisions, as a hard
process, increases with collision energy much faster than that of
light quarks. At sufficiently high energy, the produced medium will
therefore contain more charm quarks than present in a QGP at ``chemical''
equilibrium. If these charm and anticharm quarks combine at the
hadronization point statistically to form charmonium states, this new
combination mechanism should lead to a much enhanced \J~production rate,
even if all primary (``direct'') \J's are dissociated.
The two predictions, sequential suppression vs.\ statistical
regeneration, thus present two really opposite patterns, and first
LHC results should be able to distinguish between them. It should be
emphasized that statistical recombination would on one hand provide
clear evidence for the presence of a thermal medium, including even
charm quarks. On the other, it presupposes a ``new'' statistical
charmonium production mechanism, quite distinct from the hard
production forms generally discussed, in which only $c\bar c$ pairs
at very short separation can bind to form a \J. 

\section{Summary}

We have seen that in strong interaction thermodynamics, there exists
at vanishing baryon density a well-defined transition, in which 
\vspace*{-0.2cm}
\begin{itemize}
\item{color deconfinement sets in and chiral symmetry is restored,}
\vspace*{-0.2cm}
\item{the energy density increases by the latent heat of deconfinement;}
\vspace*{-0.2cm}
\item{the critical temperature $T_c$ is about $175 \pm 10$ MeV.}
\vspace*{-0.1cm}
\end{itemize}
For $T > T_c$, the state of matter is a plasma of deconfined quarks and
gluons, which can be probed by
\vspace*{-0.2cm}
\begin{itemize}
\item{electromagnetic radiation,}
\vspace*{-0.2cm}
\item{quarkonium spectra,}
\vspace*{-0.2cm}
\item{jet quenching.}
\vspace*{-0.1cm}
\end{itemize}
A more extensive survey of QCD thermodynamics will appear soon 
\cite{Springer}. -- Finally, we have given three rather model-independent 
bench-marks for a QGP 
study through very high energy nuclear collisions, 
\vspace*{-0.2cm}
\begin{itemize}
\item{does the source size (finally) increase with collision energy?}
\vspace*{-0.2cm}
\item{does the thermal photon
temperature increase with collision energy?}
\vspace*{-0.2cm}
\item{does quarkonium production show sequential suppression or
statistical regeneration?}
\end{itemize}
\vspace*{-0.1cm}
The LHC should be able to provide some first answers to these question within
the first year of heavy ion operation.

\end{document}